\documentclass[final,5p,times]{elsarticle}
\usepackage{eurosym}
\usepackage{amsmath}
\usepackage{amssymb}
\usepackage{graphicx}
\usepackage{subfigure}
\usepackage[hypertex]{hyperref}
\usepackage{amsfonts}
\usepackage{amssymb}
\usepackage{natbib}
\usepackage[usenames]{color}
\usepackage{color}
\usepackage[T1]{fontenc}
\usepackage[english]{babel}
\usepackage{float}

\hypersetup{
  colorlinks=true,linkcolor=blue,citecolor=blue,
  filecolor=blue,urlcolor=blue,breaklinks=true
}
\biboptions{sort&compress}
\journal{Physics Letters A}
\begin{document}

\begin{frontmatter}

\title{Influence of a Pseudo Aharonov-Bohm field on the quantum Hall effect in Graphene}

\author[ufla]{Cleverson Filgueiras}
\ead{cleverson.filgueiras@dfi.ufla.br}
\author[ufcg]{Rosinildo Fidelis}
\author[cftp,ufcg]{Diego Cogollo}
\ead{diegocogollo@gmail.com}
\author[ufma]{Edilberto O. Silva}
\ead{edilbertoo@gmail.com}

\address[ufla]{Departamento de F\'{i}sica, Universidade Federal de Lavras, Caixa Postal 3037, 37200-000, Lavras-MG, Brazil}
\address[cftp]{CFTP, Departamento de F\'{i}sica, Instituto Superior T\'{e}cnico, Universidade de Lisboa, Avenida Rovisco Pais 1, 1049 Lisboa, Portugal}
\address[ufcg]{Unidade Acadêmica de F\'{i}sica, Universidade Federal de Campina Grande, POB
10071, 58109-970, Campina Grande-PB, Brazil}
\address[ufma]{Departamento de F\'{i}sica, Universidade Federal do Maranh\~{a}o,
  65085-580, S\~{a}o Lu\'{i}s-MA, Brazil}

\date{\today}

\begin{abstract}
The effect of an Aharonov-Bohm (AB) pseudo magnetic field on a two
dimensional electron gas in graphene is investigated. We consider it modeled
as in the usual AB effect but since such pseudo field is supposed to be
induced by elastic deformations, the quantization of the field flux is
abandoned. For certain constraints on the orbital angular momentum
eigenvalues allowed for the system, we can observe the zero Landau level
failing to develop, due to the degeneracy related to the Dirac valleys $K$
and $K^\prime$ which is broken. For integer values of the pseudo AB flux,
the actual quantum Hall effect is preserved. Obtaining the Hall conductivity
by summing over all orbital angular momentum eigenvalues, the zero Landau
levels is recovered. Since our problem is closed related to the case where
topological defects on a graphene sheet are present, the questions posed
here are helpful if one is interested to probe the effects of a singular
curvature in these systems.
\end{abstract}

\begin{keyword}
Landau Levels \sep Hall conductivity \sep Graphene \sep Elastic deformations
\end{keyword}

\end{frontmatter}

\section{Introduction}

The study of quantum dynamics for particles in constant magnetic \cite
{Book.1981.Landau} and Aharonov-Bohm (AB) flux fields \cite{PR.1959.115.485}, which are perpendicular to the plane where the particles are confined, has
been carried out over the last years. The existence of other potentials are
also included, depending on the purpose of the investigation. For example,
in Ref. \cite{SST.1996.11.1635} an exactly soluble model to describe quantum
dots, antidots, one-dimensional rings and straight two-dimensional wires in
the presence of such fields was proposed. It is an ideal tool to investigate
the AB effects and the persistent currents in quantum rings, for instance.
In Ref. \cite{AoP.2015.353.283}, the exact bound-state energy eigenvalues and the
corresponding eigenfunctions for several diatomic molecular systems in a
pseudoharmonic potential were analytically calculated for any arbitrary
angular momentum. The Dirac bound states of anharmonic oscillator under
external magnetic and AB flux fields were addressed recently \cite
{AoP.2014.341.153}. The investigation of a Cornell Potential in external
fields was considered in Ref. \cite{FBS.2014.55.1055}. Other examples can be
found elsewhere. It would be interesting to carry out such investigations on
graphene, an one atom thick material which rapidly caught the attention of
many physicists \cite{NatureMat.2007.6.183}. Graphene, a single layer of
carbon atoms in a honeycomb lattice, is considered a truly two dimensional
system. The carriers within it behave as two-dimensional massless Dirac
fermions \cite{RMP.2009.81.109}. Due to its peculiar physical properties,
graphene has great potential for nanoelectronic applications \cite
{NL.2008.8.902,Schwierz2010,LiChengShashurinEtAl2012}. Graphene can be
considered a zero-gap semiconductor. This fact prevent the pinch off of
charge currents in electronic devices. Quantum confinement of electrons and
holes in nanoribbons \cite{SolsGuineaNeto2007} and quantum dots \cite
{HanOezyilmazZhangEtAl2007} can be realized in order to induce a gap.
However, this lattice disorder suppress an efficient charge transport \cite
{MuccioloCastroNetoLewenkopf2009}. One alternative to open a gap is to
induce a strain field in a graphene sheet onto appropriate substrates \cite
{MuccioloCastroNetoLewenkopf2009}. They play the role of an effective gauge
field which yields a pseudo magnetic field \cite
{VozmedianoKatsnelsonGuinea2010}. Unlike actual magnetic fields, these
strain induced pseudo magnetic fields do not violate the time reversal
symmetry \cite{GuineaKatsnelsonGeim2010}. A numerical study on the
uniformity of the pseudomagnetic field in graphene as the relative
orientation between the graphene lattice and straining directions was
carried out in Ref. \cite{VerbiestBrinkerStampfer2015} and it was observed that
observing the pseudomagnetic field in Raman spectroscopy setup is feasible.
In Ref. \cite{VerbiestStampferHuberEtAl2016}, two different mechanisms that could
underlie nanometer-scale strain variations in graphene as a function of
externally applied tensile strain is presented.

Recently, some works devoted to the search for solutions of the Dirac
equation with position dependent magnetic fields were addressed in this context \cite
{SetareOlfati2007,KuruNegroNieto2009,HartmannPortnoi2014,EshghiMehraban2016,Eshghi2016}. However, they consider \textit{actual} instead of \textit{pseudo} magnetic
fields. No experiments have been reported yet and we believe it is because
such field configurations are not easy to implement in the laboratory. In
this paper, we investigate a graphene sheet in the presence of both a
constant orthogonal magnetic and a pseudo Aharonov-Bohm field. This kind of
pseudo AB field was first investigated in \cite
{JuanCortijoVozmedianoEtAl2011}, where a device to detect microstresses in
graphene able to measure AB interferences at the nanometer scale was
proposed. It was showed on it that fictitious magnetic field associated with
elastic deformations of the sample yield interferences in the local density
of states. Here, we consider such pseudo AB-field modeled like a
\textquotedblleft thin solenoid\textquotedblright\ so we can analytically
investigate the consequences in the Quantum Hall effect, for instance.
Specifically, we investigate how such non constant pseudo AB field modify
the relativistic Landau levels and, as consequence, the Hall conductivity in
a suspended graphene. Since such pseudo AB field is modeled as the actual AB
field, which is yielded by a thin tube flux, either regular or irregular
wavefunctions can be solution of the problem. This is in agreement with
other quantum problems where singularities have also appeared. This question
about the correct behavior of wavefuntion whenever we have singularities has
been investigated via the \textit{self adjoint extension approach} over the
last years \cite
{AoP.2010.325.2529,JMP.2012.53.122106,PLB.2013.719.467,AoP.2013.339.510,PRD.2012.85.041701,TMP.2010.163.511}
. In our case, if singular effects manifest, then a constraint in the
orbital angular momentum eigenvalues appears.

An important result is that, for certain constraints on the orbital angular
momentum eigenvalues allowed for the system, the zero-energy, which exist in
the known relativistic Landau levels when just the constant orthogonal
magnetic field is present, does not develop around both valleys, $K$ and $
K^{\prime}$. The consequence is that a Hall plateau develop at the null
\textit{filling factor} (dimensionless ratio between the number of charge
carries and the flux quanta). This is due to the degeneracy related to the
Dirac valleys $K$ and $K^\prime$ which is broken. For the integer values of
the AB flux, the zero energy manifests again around both valleys. On the
other hand, by analyzing the quantum Hall conductivity summing for all the
orbital angular momentum eigenvalues allowed for the system, we observe that
the standard plateaux at all integer $n$ of $2e^2/h$ will show up, including
that for $n=0$. This is also due to the degeneracy related to the Dirac
valleys $K$ and $K^\prime$ which is broken for certain part of the energy
spectrum. This is in contrast to the usual Quantum Hall effect in graphene,
where the quantum Hall conductivity exhibits the standard plateaux at all
integer $n$ of $4e^2/h$, for $n=1,2,3..$, and $2e^2/h$ for $n=0$.

The plan of this work is the following. First, we investigate how a varying
pseudo magnetic field perpendicular to a graphene sheet is going to affect
the relativistic Landau levels. Then, we investigate the influence of such
pseudo AB field in the quantized Hall conductivity. At the end, we have the
concluding remarks.

\section{Modifications in the Relativistic Landau Levels under pseudo-AB
field}

In this section, we will investigate how a pseudo AB field is going to affect the relativistic Landau
levels. First, we must remember the reader that the low energy excitations
of graphene behave as massless Dirac fermions, instead of massive electrons.
Their internal degrees of freedom are: sublattice index (pseudospin), valley
index (flavor) and real spin, each taking two values. The real spin is
irrelevant in our problem and will not be taken into account, except for an
additional degeneracy factor $2$ in the Hall conductivity. Then, the low
energy excitations around a valley is described by the $(2+1)-$dimensional
Dirac equation as
\begin{equation}
-i\hbar v_{F}\left( \mathbf{\sigma }\cdot \mathbf{\nabla }\right) \Psi (\mathbf{r}
)=E\Psi (\mathbf{r}),  \label{dirac}
\end{equation}
where $\sigma =\left( \sigma _{x},\sigma _{y}\right) $ are the Pauli
matrices, $\Psi =(\varphi _{1},\varphi _{2})^{T}$ is a two component spinor
field, the speed of light $c$ was replaced by the Fermi velocity ($
v_{F}\approx 10^{6}$m/s) and $\hbar $ has been fixed equal to one. The
electronic states around the zero energy are states belonging to distinct
sublattices. This is the reason we have a two component wavefunction. Two
indexes to indicate these sublattices, similar to spin indexes (up and down),
must be used. The inequivalent corners of the Brillouin zone, which are
called \textit{Dirac points}, are labeled as $K$ and $K^{^{\prime }}$(valley
index) \cite{RMP.2009.81.109}. In what follows, we consider $v_F=\hbar=1$. We reinstate the proper
units latter, in the analysis of our main results.

In this work, the pseudo varying magnetic field is supposed to appear due
strains on a graphene sheet \cite{VozmedianoKatsnelsonGuinea2010}. The
valleys $K$ and $K^{^{\prime }}$ feel an effective field of $\tilde{\mathbf{A
}}\pm \mathbf{A}$, where $\tilde{\mathbf{A}}$ is due to a real magnetic
field and $\mathbf{A}$ is due to a pseudo-magnetic field. Notice that a
different sign has to be used for the gauge field due to strain at the
valleys $K$ and $K^{^{\prime }}$ since such fields do not break time
reversal symmetry \cite{GuineaKatsnelsonGeim2010}. These vector potentials
can be inserted into the Dirac equation via a minimal coupling, $\mathbf{p}
\rightarrow \mathbf{p}-e\mathbf{A}$. We begin by writing the massless Dirac
equation for the four-component spinor $\Psi $
\begin{equation}
\left[ \beta \mathbf{\gamma }\cdot \left( \mathbf{p}-e\mathbf{A}
\right) -E\right] \Psi \left( \mathbf{r}\right) =0,  \label{diracA}
\end{equation}
with the $\beta ,\mathbf{\gamma }$ matrices being given in terms of the
Pauli matrices as \cite{NPB.1989.328.140}
\begin{equation}
\beta =\sigma ^{z},\qquad \beta \gamma ^{1}=\sigma ^{1},\qquad \beta \gamma
^{2}=s\sigma ^{2},  \label{newmatrices}
\end{equation}
where the parameter $s$, which has a value of twice the spin value, can be
introduced to characterizing the two pseudo spin states, with $s=+1$ for spin
\textquotedblleft up\textquotedblright\ and $s=-1$ for spin
\textquotedblleft down\textquotedblright. Equation (\ref{diracA}) can be
placed on a quadratic form by applying the operator $E+\beta \mathbf{\gamma }
\cdot \left( \mathbf{p}-e\mathbf{A}\right) $. The result of this application
provides the Dirac-Pauli equation
\begin{equation}
\left[ \mathbf{p}^{2}-2e\left( \mathbf{A}\cdot \mathbf{p}\right) +
e^{2}\mathbf{A}^{2}-es\hbar\left( \mathbf{\sigma }
\cdot \mathbf{B}\right) \right] \psi \left( \mathbf{r}\right) =
E^{2}\psi\left( \mathbf{r}\right) ,  \label{pauliA}
\end{equation}
where $\psi \left( \mathbf{r}\right) $ is a a four-component spinorial wave
function. We consider the case where the particle interacts with a gauge
field
\begin{equation}
\mathbf{A=A}_{1}+\mathbf{A}_{2},  \label{vectorA}
\end{equation}
with
\begin{equation}
\mathbf{A}_{1}=\frac{B_{0}r}{2}\mathbf{\hat{\varphi}},~~~\mathbf{A}_{2}=
\frac{\phi }{r}\mathbf{\hat{\varphi}},  \label{potential}
\end{equation}
where $B$ is the magnetic field magnitude and $\phi $ is the flux
parameter. The potentials in Eq. (\ref{vectorA}) both provide one magnetic
field perpendicular to the plane $\left( r,\varphi \right) $, namely
\begin{equation}
\mathbf{B}=\mathbf{B}_{1}+\mathbf{B}_{2},  \label{fieldB}
\end{equation}
with
\begin{eqnarray}
\mathbf{B}_{1} &=&\mathbf{\nabla }\times \mathbf{A}_{1}=B\mathbf{\hat{z}}
,  \label{fieldB1} \\
\mathbf{B}_{2} &=&\mathbf{\nabla }\times \mathbf{A}_{2}=\phi \frac{\delta
\left( r\right) }{r}\mathbf{\hat{z}}.  \label{fieldB2}
\end{eqnarray}
Note that the field (\ref{fieldB2}) is one produced by a solenoid. If the
solenoid is extremely long, the field inside is uniform, and the field
outside is zero. But, the vector potential outside the solenoid is not zero.
However, in a general dynamics, the particle is allowed to access the $r=0$
region. In this region, the pseudo magnetic field is non-null. If the radius of the
solenoid is $r_{0}\approx 0$, then the relevant magnetic field is $\mathbf{B}_{2}\sim \delta \left( r\right) $ as in Eq. (\ref{fieldB2}).

Adopting the decomposition
\begin{equation}
\left(
\begin{array}{c}
\psi _{1}\left( r,\varphi \right)  \\
\psi _{2}\left( r,\varphi \right)
\end{array}
\right) =\left(
\begin{array}{c}
\sum\limits_{m}f_{m}\left( r\right) \;e^{im\varphi } \\
i\sum\limits_{m}g_{m}\left( r\right) \;e^{i(m+s)\varphi }
\end{array}
\right) ,  \label{ansatz}
\end{equation}
with $m+1/2=\pm 1/2,\pm 3/2,\ldots $, with $m\in \mathbb{Z}$, and inserting
this into Eq. (\ref{pauliA}), the equation for $f_{m}\left( r\right) $ is
found to be
\begin{equation}
hf_{m}\left( r\right) =k^{2}f_{m}\left( r\right),  \label{diffA}
\end{equation}
where
\begin{equation}
k^{2}=E^{2}+\left(m+s\right)eB-e^{2}\phi B,
\end{equation}
\begin{equation}
h=h_{0}-se\phi \frac{\delta \left( r\right) }{r},
\label{hfull}
\end{equation}
\begin{equation}
h_{0}=-\frac{d^{2}}{dr^{2}}-\frac{1}{r}\frac{d}{dr}+\frac{1}{r^{2}}\left(
m-\lambda \right) ^{2}+\omega ^{2}r^{2},
\end{equation}
and $\lambda =e\phi$ and $\omega ={eB}/{2}$. Note that Eq. (\ref{diffA}) contains the $
\delta $ function in the radial Hamiltonian $h$, which is singular at the
origin. In order to deal with a Hamiltonian of this nature we making use of
the self-adjoint extension approach \cite
{Book.2004.Albeverio,JMP.1985.26.2520}. According to Ref. \cite
{Book.1975.Reed.II}, the Hamiltonian $h_{0}$ is essentially self-adjoint if $
\left\vert m-\lambda \right\vert \geq 1$, while for $\left\vert m-\lambda
\right\vert <1$ it admits an one-parameter family of self-adjoint
extensions, $h_{0,\zeta _{m}}$, where $\zeta _{m}$ is the self-adjoint
extension parameter. To characterize this family of self-adjoint extensions,
we use the approach proposed in \cite{JMP.1985.26.2520}, which uses the
boundary condition at the origin
\begin{equation}
f_{0}=\zeta _{m}f_{1},  \label{bc}
\end{equation}
with
\begin{align*}
f_{0}={}& \lim_{r\rightarrow 0^{+}}r^{|m-\lambda |}f_{m}\left( r\right) , \\
f_{1}={}& \lim_{r\rightarrow 0^{+}}\frac{1}{r^{|m-\lambda |}}\left[
f_{m}\left( r\right) -f_{0}\frac{1}{r^{|m-\lambda |}}\right] ,
\end{align*}
where $\zeta _{m}\in \mathbb{R}$ is the self-adjoint extension parameter. In
the boundary condition above, if\ $\zeta _{m}=0$, we have the free
Hamiltonian (without the $\delta $ function) with regular wave functions at
the origin, and for $\zeta _{m}\neq 0$, the boundary condition in Eq. (\ref
{bc}) permit an $r^{-\left\vert m-\lambda \right\vert }$ singularity in the
wave functions at the origin.

\section{The bound state energy and wave function}

With the application of the boundary condition (\ref{bc}), we can find the
energy spectrum of the system. Before doing this, first we make a variable
change in Eq. (\ref{diffA}), $\rho =\omega r^{2}$, so that it is written as (
$r\neq 0$)
\begin{equation}
\rho \frac{d^{2}f_{m}}{d\rho ^{2}}+\frac{df_{m}}{d\rho }-\left[ \frac{\left(
m-\lambda \right) ^{2}}{4\rho }+\frac{\rho }{4}-\frac{k^{2}}{4\omega }\right]
f_{m}=0.  \label{edofrho}
\end{equation}
Moreover, because of the boundary condition (\ref{bc}), we seek for regular
and irregular solutions for Eq. (\ref{edofrho}). So, after studying the
asymptotic limits of Eq. (\ref{edofrho}), we find the following regular ($+$
) (irregular ($-$)) solution:
\begin{equation}
f_{m}\left( \rho \right) =\rho ^{\pm \frac{1}{2}\left\vert m-\lambda
\right\vert }e^{-\rho /2}F\left( \rho \right) .  \label{frho}
\end{equation}
Insertion of this solution into Eq. (\ref{edofrho}) yields
\begin{equation}
\rho F^{\prime \prime }\left( \rho \right) +\left( 1\pm \left\vert
m-\lambda \right\vert -\rho \right) F^{\prime }\left( \rho \right)-\left( \frac{1\pm \left\vert m-\lambda \right\vert }{2}-\frac{k^{2}}{
4\omega }\right) F\left( \rho \right) =0.  \label{edoMrho}
\end{equation}
The general solution to this equation is given in terms of the confluent
hypergeometric function of the first kind \cite{Book.1972.Abramowitz},
\begin{align}
f_{m}\left( \rho \right) & =a_{m}\rho ^{\frac{1}{2}\left\vert m-\lambda
\right\vert }e^{-\frac{\rho }{2}}\;F\left( d_{+},1+\left\vert m-\lambda
\right\vert ,\rho \right)   \notag \\
& +b_{m}\rho ^{-\frac{1}{2}\left\vert m-\lambda \right\vert }e^{-\frac{\rho
}{2}}\;F\left( d_{-},1-\left\vert m-\lambda \right\vert ,\rho \right) ,
\label{general_sol_2_HO}
\end{align}
with
\begin{equation}
d_{\pm }=\frac{1\pm \left\vert m-\lambda \right\vert }{2}-\frac{k^{2}}{
4\omega },
\end{equation}
where $a_{m}$ and $b_{m}$ are, respectively, the coefficients of the regular
and irregular solutions.

Now, by applying the boundary condition (\ref{bc}), one finds the following
relation between the coefficients $a_{m}$ and $b_{m}$
\begin{equation}
\zeta _{m}\omega ^{\left\vert m-\lambda \right\vert }=\frac{b_{m}}{a_{m}}
\left[ 1+\frac{\zeta _{m} k^{2}}{4\left( 1-\left\vert m-\lambda
\right\vert \right) }\lim_{r\rightarrow 0^{+}}r^{2-2\left\vert m-\lambda
\right\vert }\right] .  \label{coefA}
\end{equation}
Note that $\lim_{r\rightarrow 0^{+}}r^{2-2\left\vert m-\lambda \right\vert }$
diverges if $\left\vert m-\lambda \right\vert \geq 1$. This condition
implies that $b_{m}$ must be zero if $\left\vert m-\lambda \right\vert \geq 1
$ and only the regular solution contributes to $f_{m}\left( \rho \right) $.
For $\left\vert m-\lambda \right\vert <1$, when the operator $H_{0}$ is not
self-adjoint, there arises a contribution of the irregular solution to $
f_{m}\left( r\right) $ \cite
{AoP.2013.339.510,TMP.2009.161.1503,EPJC.2014.74.2708}. In this manner, the
contribution of the irregular solution for the system wave function stems
from the fact that the operator $H_{0}$ is not self-adjoint.

For $f_{m}(r)$ be a bound state wave function, it must vanish at large
values of $r$, i.e., it must be normalizable. So, from the asymptotic
representation of the confluent hypergeometric function, the normalizability
condition is translated in
\begin{equation}
\frac{b_{m}}{a_{m}}=-\frac{\Gamma \left( 1+\left\vert m-\lambda \right\vert
\right)}{\Gamma \left( 1-\left\vert m-\lambda \right\vert \right) }\frac{
\Gamma \left( d_{-}\right) }{\Gamma \left( d_{+}\right)}.  \label{coefB}
\end{equation}
From Eq. (\ref{coefA}), for $\left\vert m-\lambda \right\vert <1$, we have
\begin{equation}
\frac{b_{m}}{a_{m}}=\zeta _{m}\omega ^{\left\vert m-\lambda \right\vert }.
\end{equation}
By combining this result with (\ref{coefB}), one finds
\begin{equation}
\frac{\Gamma \left( d_{+}\right) }{\Gamma \left( d_{-}\right) }=-\frac{1}{
\zeta _{m}\omega ^{\left\vert m-\lambda \right\vert }}\frac{\Gamma \left(
1+\left\vert m-\lambda \right\vert \right) }{\Gamma \left( 1-\left\vert
m-\lambda \right\vert \right) }.  \label{energyBG}
\end{equation}
Equation (\ref{energyBG}) implicitly determines the energy spectrum for different values of
the self-adjoint extension parameter. Two limiting values for the
self-adjoint extension parameter deserve some attention. For $\zeta _{m}=0$,
when the $\delta $ interaction is absent, only the regular solution
contributes for the bound state wave function. On the other side, for $\zeta
_{m}=\infty $ only the irregular solution contribute for the bound state
wave function. For all other values of the self-adjoint extension parameter,
both regular and irregular solutions contributes for the bound state wave
function. The energies for the limiting values are obtained from the poles
of the gamma function, namely,
\begin{equation}
\left\{
\begin{array}{lll}
d_{+}=-n & \mbox{for }\zeta _{m}=0 & \mbox{(regular solution)}, \\
d_{-}=-n & \mbox{for }\zeta _{m}=\infty  & \mbox{(irregular solution)},
\end{array}
\right.   \label{poles}
\end{equation}
with $n$ a nonnegative integer, $n=0,1,2,\ldots $. By manipulation of Eq. (
\ref{poles}), we obtain
\begin{equation}
E^{2}=eB\left[ 2n+1\pm \left\vert m-\lambda \right\vert +\lambda
-m-s\right] .  \label{energy}
\end{equation}
In particular, it should be noted that for the case when $\left\vert
m-\lambda \right\vert \geq 1$ or when the $\delta $ interaction is absent,
only the regular solution contributes for the bound state wave function ($
b_{m}=0$), and the energy is given by Eq. (\ref{energy}) with plus sign. The
energy spectrum above must be analyzed in terms of the values that the
parameter $\lambda $ can assume, since the condition $\left\vert m-\lambda
\right\vert \geq 1$($\left\vert m-\lambda \right\vert < 1$) for the
regular(irregular) solution has to be fulfilled. Let us consider the regular
wave functions at first. Then, the spectrum (\ref{energy}) in this case can
be written as
\begin{equation}
E_{n}^{+}=\pm \sqrt{2eBn^{\prime }}  \label{energy1}
\end{equation}
for $m-\lambda >1$, with $n^{\prime }=n+\frac{1-s}{2}=0,1,2,3,...$, and
\begin{equation}
E_{n}^{-}=\pm \sqrt{2eB\left( n^{\prime }+\lambda \right) },  \label{energy2}
\end{equation}
for $m-\lambda <-1$, with $n^{\prime }=n+\frac{1-s}{2}-m=0,1,2,3,...$. The
energy spectrum around the Dirac point $K^{\prime }$ is obtained by changing
$\lambda \rightarrow -\lambda $. This way, we have
\begin{equation}
E_{n}^{+}=\pm \sqrt{2eBn^{\prime }}  \label{energy3}
\end{equation}
for $m-\lambda >1$, with $n^{\prime }=n+\frac{1-s}{2}=0,1,2,3,...$, and
\begin{equation}
E_{n}^{-}=\pm \sqrt{2eB\left( n^{\prime }+1-\lambda \right) }.
\label{energy4}
\end{equation}
for $m-\lambda <-1$, with $n^{\prime }=n+\frac{1-s}{2}-m=0,1,2,3,...$. If we
had chosen $\lambda <0$ at first, the results above are the same, but
exchanged between $K^{\prime }$ and $K$. If $\lambda $ is integer, the
energy spectrum of the system will be the same as in the case without such
AB field, $E_{n}^{-}=\pm \sqrt{2eBn}$, since $n=n^{\prime }\pm \lambda
=0,1,2,3,...$. In this case, the parameter $\lambda $ does not splits the
energy levels and the degeneracy regardless the Dirac valleys is not broken.

\begin{figure}[H]
\center
\subfigure[]{\includegraphics[width=8.0cm,height=7.0cm]{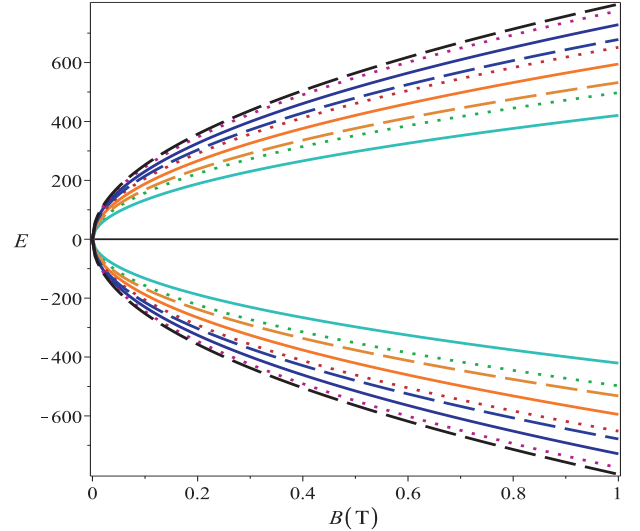}}
\qquad
\subfigure[]{\includegraphics[width=8.0cm,height=7.0cm]{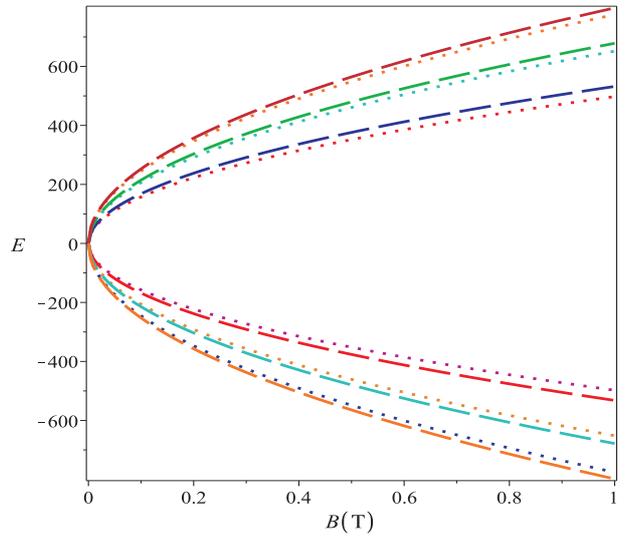}}
\caption{Plot of the energy versus the magnetic field, for $\protect\lambda
=0.4$; In Fig. (a), any value of the angular momentum quantum number $m$
being possible. The zero energy is present. In Fig. (b), we have the case
for the allowed values of $m$ for which only the part of the energy spectrum
containing the parameter $\protect\lambda $ is present. The zero energy is
absent.}
\label{EnergyMag}
\end{figure}

We now turn our attention to the case considering the irregular solution,
with the condition $\rvert m-\lambda \lvert <1$. Remember that we now take
into account the minus sign in Eq. (\ref{energy}). The energy spectrum will
be written in the same way as Eqs. (\ref{energy1}), (\ref{energy2}), (\ref
{energy3}) and (\ref{energy4}), but the constraint $\rvert m-\lambda \lvert
<1$ will permit only some values of allowed $m$, that is, $-1+\lambda
<m<1+\lambda $. Moreover, Eqs. (\ref{energy1}) and (\ref{energy3}) hold for $
-1<m-\lambda \leq 0$, while Eqs. (\ref{energy2}) and (\ref{energy4}) hold
for $0<m-\lambda <1$. The energy levels are depicted in Fig. \ref{EnergyMag}.

\section{The effect of AB pseudo field on the Hall conductivity}

In this section, we investigate the influence of such AB field in the
quantized Hall conductivity. We express the energy scale associated with the
magnetic field in the units of temperature. This way, we have
\begin{eqnarray}
eB &\rightarrow &\frac{eBv_{F}^{2}}{c}=\frac{eB\hbar v_{F}^{2}}{c}\frac{1}{
k_{B}^{2}}(K^{2})  \notag \\
&=&8.85\times 10^{-8}v_{F}^{2}(m/s)B\mathrm{(T)},
\end{eqnarray}
where $v_{F}$ and $B$ are given in $m/s$ and Tesla, respectively.

In the last section, we have found the energy levels using the polar
coordinates but we consider that the sample is not a disc, that is, it has a
rectangular shape. This way, we start by considering the expression for the
Hall conductivity obtained in Ref. \cite{GusyninSharapov2006a} in the clean
limit, that is,
\begin{align}
\sigma _{xy}^{(1)}& =-\frac{e^{2}N_{f}\mathrm{sign}(eB)}{4\pi }  \notag \\
& \times \sum_{n=0}^{\infty }\alpha _{n}\left[ \tanh \left( \frac{\mu +E_{n}
}{T}\right) +\tanh \left( \frac{\mu -E_{n}}{T}\right) \right],
\end{align}
where $\alpha _{0}=1$ (for $n=0$) and $\alpha _{n}/\alpha _{0}=2$ (for $n\geq
1 $).

\begin{figure}[h!]
\center
\subfigure[]{\includegraphics[width=8.0cm,height=7.0cm]
{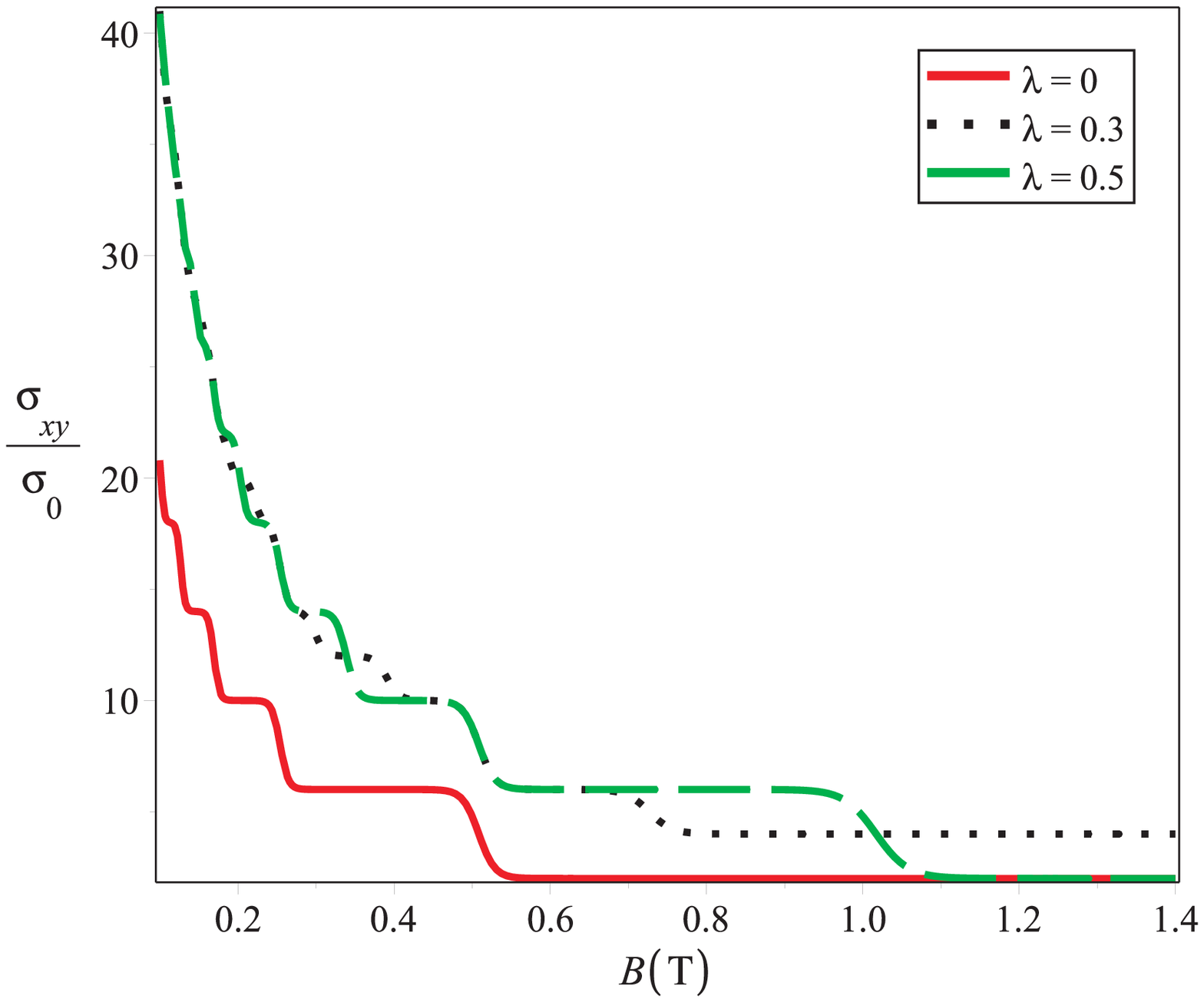}}
\qquad
\subfigure[]{\includegraphics[width=8.0cm,height=7.0cm]
{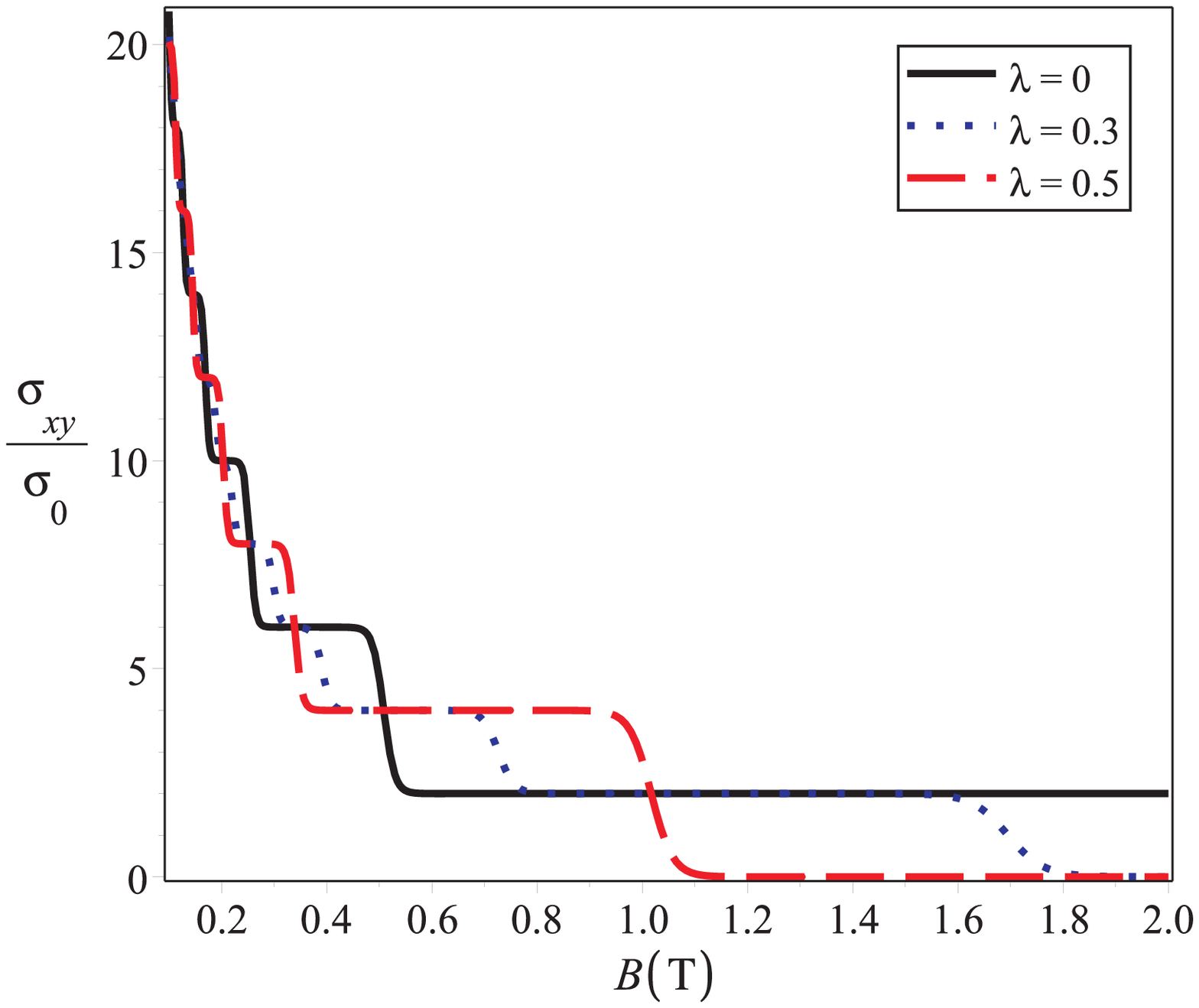}}
\caption{Plot of the Hall conductivity versus the magnetic field for $T=3K$:
$\protect\lambda =0$ (absence of the elastic interaction), $\protect\lambda
=0.3$ and $\protect\lambda =0.5$; In Fig. (a), any value of the angular
momentum quantum number $m$ being possible (see Fig. \ref{EnergyMag}(a)). The plateaux
shift to higher magnetic fields. In Fig. (b), we have the case for the
allowed values of $m$ for which only the part of the energy spectrum
containing the parameter $\protect\lambda $ is present (see Fig. \ref{EnergyMag}(b)).}
\label{HallMag}
\end{figure}

\begin{figure}[H]
\center
\subfigure[]{\includegraphics[width=8.0cm,height=7.0cm]{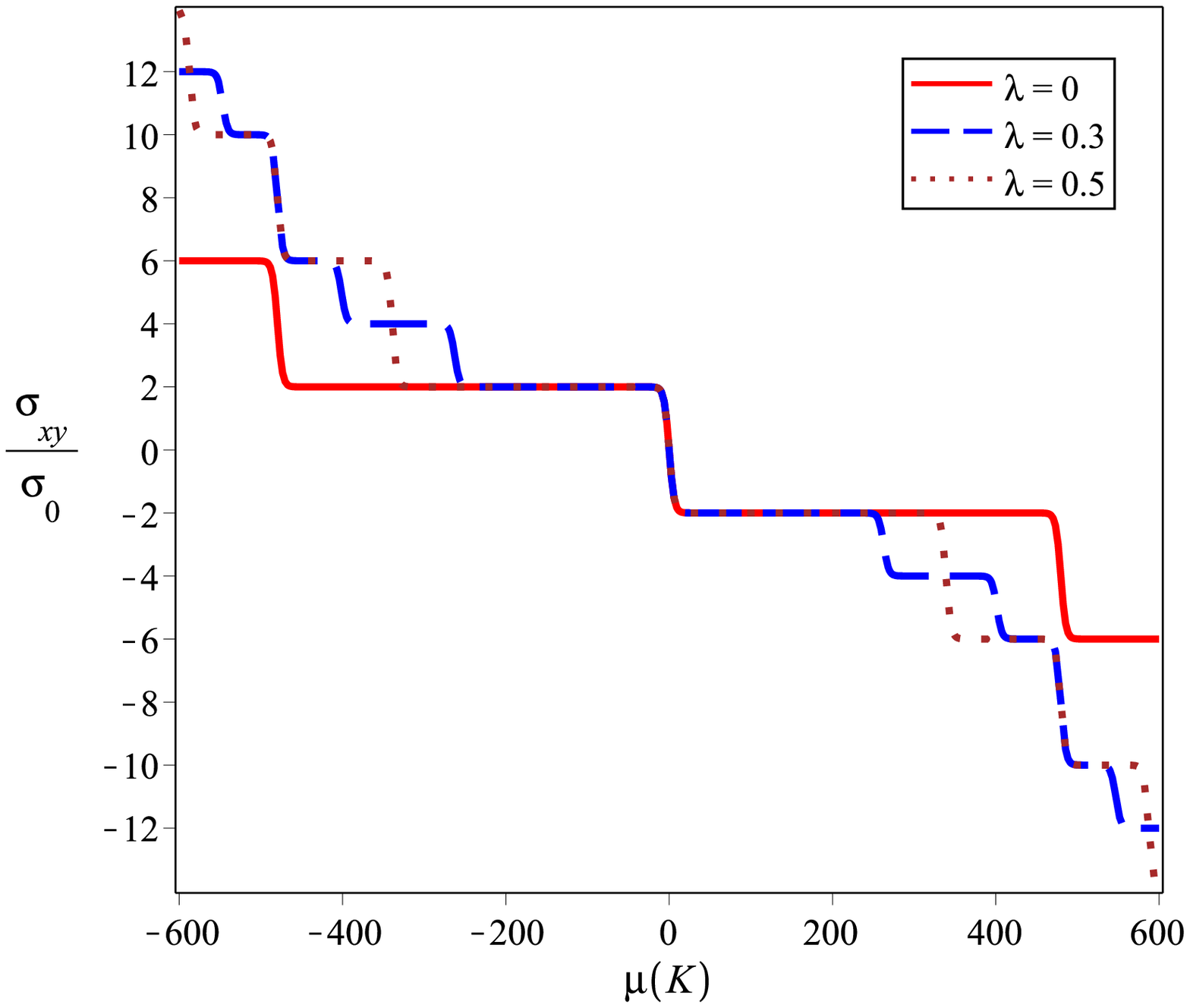}}
\qquad
\subfigure[]{\includegraphics[width=8.0cm,height=7.0cm]{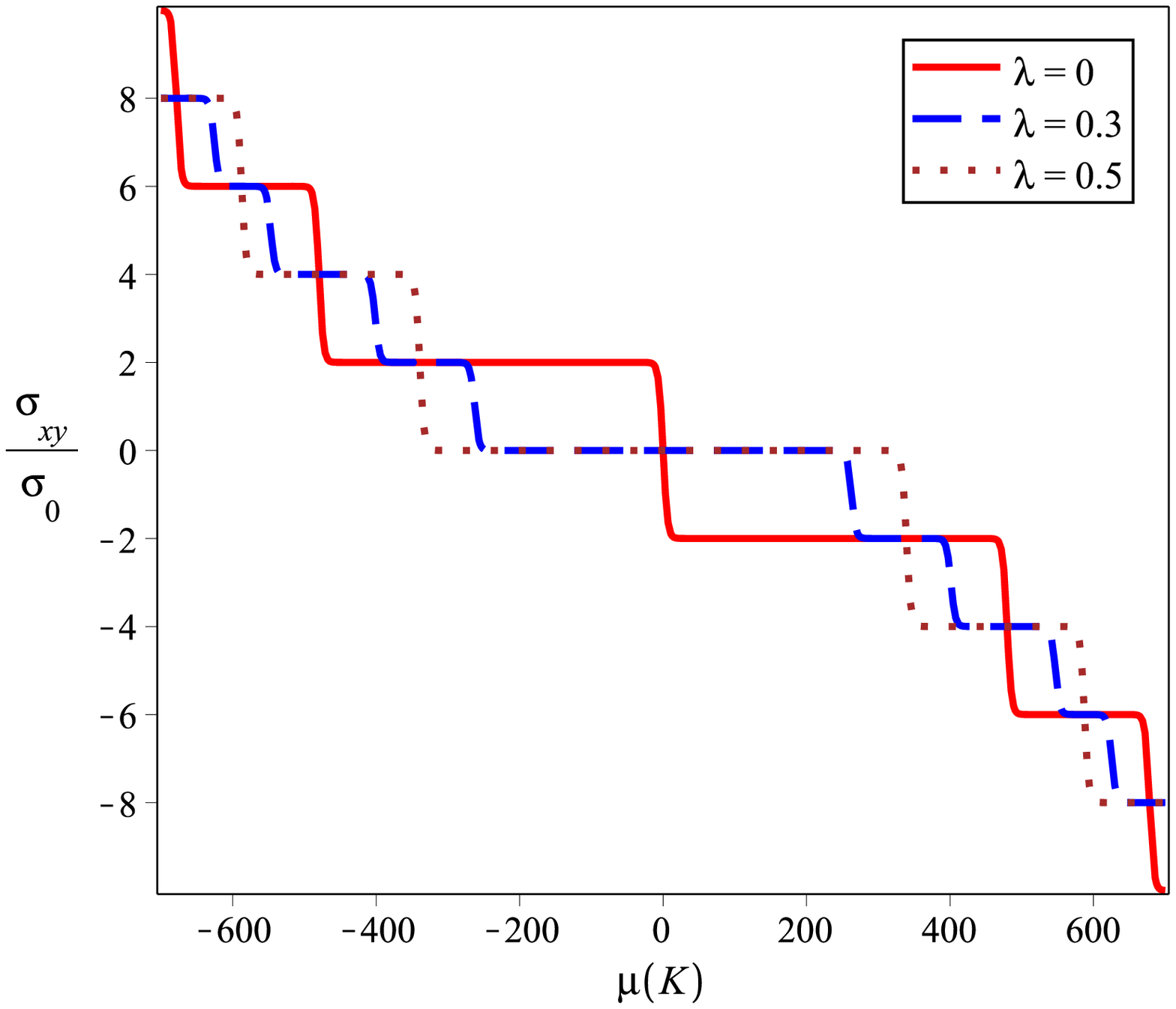}}
\caption{Plot of the Hall conductivity versus the chemical potential $
\protect\mu $ for $T=3K$, $\protect\lambda =0$ (absence of the elastic
interaction), $\protect\lambda =0.3$ and $\protect\lambda =0.5$; In Fig. (a),
any value of the angular momentum quantum number $m$ being possible (see
Fig. \ref{EnergyMag}(a) and Fig. \ref{HallMag}(a)). In Fig. (b), we have the case for the allowed values of
$m$ for which only the part of the energy spectrum containing the parameter $
\protect\lambda $ is present (see Fig. \ref{EnergyMag}(b) and Fig. \ref{HallMag}(b)). In this case, a
plateau develops at a null filling factor.}
\label{HallCh}
\end{figure}

This is related to the above-mentioned smaller degeneracy of the $n=0$
Landau level. In our case, we have observed that such AB elastic field, with
non integer $\lambda$, fails to observe the zero Landau level around one
valley. Then, the degeneracy of energy levels related to these valleys is
broken due to the pseudo AB field in this situation. Moreover, the energy
levels are shifted differently around the each valley. The consequence is
that we have to consider a sum in the valley index. Therefore, we have the
Hall conductivity as,
\begin{align}
\sigma _{xy}&=\sigma _{xy}^{(1)}-\frac{e^{2}N_{f}\mathrm{sign}(eB)}{4\pi }
\sum_{k=K}^{K^{\prime }}\sum_{n^{\prime}=0}^{\infty }\alpha
_{n^{\prime}}\left(\lambda\right)  \notag \\
&\times\left[ \tanh \left( \frac{\mu +E^{k}_{n^{\prime}} \left(\lambda\right)
}{T}\right) +\tanh \left( \frac{\mu -E^{k}_{n^{\prime}}\left(\lambda\right)}{
T}\right) \right],
\end{align}
where $\alpha _{n^{\prime}}\left(\lambda\right)=1,$ for any $n^{\prime}$.
\begin{figure}[h]
\includegraphics[width=8.0cm,height=7.0cm]{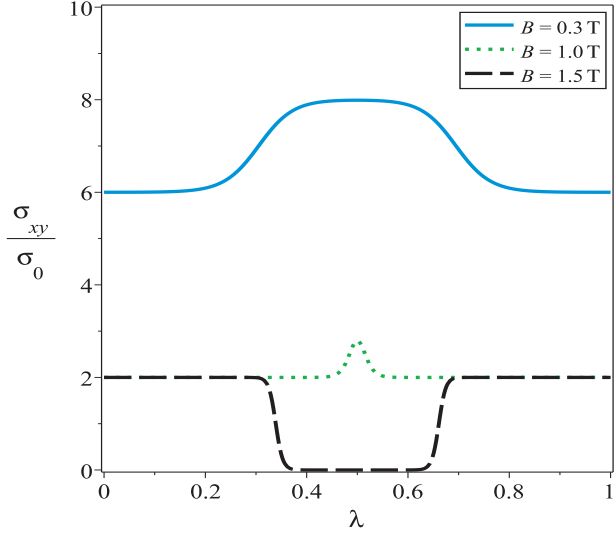}
\caption{Plot of Hall conductivity versus the elastic parameter $\protect
\lambda $ for $T=3\mathrm{K}$ (we ignore the terms which do not contain $
\protect\lambda $). Notice that it is symmetric around $\protect\lambda =0.5$
}
\label{Hall.Elastic}
\end{figure}
In Fig. \ref{Hall.Elastic}, we plot the Hall conductivity versus the elastic
parameter $\lambda $, which shows that the Hall conductivity is symmetric
around $\lambda =0.5$. In Fig. \ref{HallMag}(a), we plot the Hall
conductivity versus the magnetic field for different values of $\lambda $.
In the Quantum Hall effect in graphene ($\lambda \equiv 0$), the quantum
Hall conductivity exhibits the standard plateaux at all integer $n$ of $
4e^{2}/h$, for $n=1,2,3..$, and $2e^{2}/h$ for $n=0$. For $\lambda \neq 0$,
the plateaux are shifted to higher magnetic fields. The presence of the
Pseudo AB field introduce intermediate plateaux between them. These extra
plateaux are not observed for $\lambda =0.5$, since the valley degeneracy is
recovered in this case. In Fig. \ref{HallCh}(a) it is depicted the Hall
conductivity versus the chemical potential for some values of $\lambda $. We
observe the same fact, that is, the quantum Hall conductivity exhibits the
standard plateaux at all integer $n$ of $2e^{2}/h$. In the Fig. \ref{HallMag}(b) and
\ref{HallCh}(b), we analyze the case supposing that the system is prepared so that only
the energies containing the parameter $\lambda $ are possible to be occupied
by the electrons. This happens if only either the condition $m-\lambda >1$
(regular solutions) or $-1<m-\lambda <0$ (irregular solutions) hold. In this
case, a Hall plateau develops at $\sigma _{xy}=0$ (filling factor $\nu =0$),
in contrast to what we have discussed above. This happens when a gap is
opened in the energy bands of graphene.

\section{Concluding Remarks}

In this work, we investigated how the relativistic Landau levels and the
quantum Hall conductivity are modified if fermions on graphene are held in
the presence of a constant orthogonal magnetic field together with a AB
pseudo-field. We considered it mathematically modeled as in the case of a
thin solenoid in the case of an actual AB field. We were able to study this
problem analytically since our squared Dirac equation yielded a differential
equation whose solutions are well established in terms of Hypergeometric
series, which appear in many contexts, helping addressing different physical
problems analytically as we did here. We have observed that, for certain
constraints on the orbital angular momentum eigenvalues allowed for the
system ($m-\lambda \geq 1$ for regular wavefunctions and $-1<m-\lambda \leq 0
$ for irregular ones), it fails to observe the zero Landau level around both
valleys, $K$ and $K^{\prime }$. The consequence is that a Hall plateau
develops at the filling factor $\nu =0$. Then, the quantum Hall conductivity
showed the standard plateaux at all integer $n$ of $2e^{2}/h$ except for $n=0
$. This is in contrast to the usual Quantum Hall effect in graphene, where
the quantum Hall conductivity exhibits the standard plateaux at all integer $
n$ of $4e^{2}/h$, for $n=1,2,3..$, and $2e^{2}/h$ for $n=0$. Without such
constraints in the orbital angular momentum eigenvalues, the the zero Landau
level around both valleys are recovered and we have the plateaux at all
integer $n$ of $2e^{2}/h$ including that for $n=0$.

As a final word, we theoretically described a way to manipulate the
relativistic Landau levels by assuming the existence of an AB pseudo field.
Graphene under different position-dependent magnetic fields was investigated
theoretically in reference \cite
{SetareOlfati2007,KuruNegroNieto2009,HartmannPortnoi2014,EshghiMehraban2016,Eshghi2016}. It would also be interesting to investigate them as pseudo magnetic fields
combined with a constant magnetic field as we have done in this work. If
either simulations or experiments involving graphene fail to observe the
zero Landau level, the presence of varying pseudo magnetic fields should be
investigated. The problem addressed here is closed related to but not equal
to the case where topological defects on a graphene sheet are present, since
their existence also split the zero energy \cite
{BuenoFurtadoM.Carvalho2012,BiswasSon2016}. Here, we have showed how a delta
like interaction can affect the quantum hall system and it is important to
have those questions in mind if one is interested to probe the effects of a
singular curvature in these systems.

\section*{Acknowledgments}

This work was supported by the Brazilian agencies CNPq, FAPEMA and FAPEMIG.

\bibliographystyle{model1a-num-names}

\end{document}